\documentclass{aastex631}

\usepackage{tikz}
\usetikzlibrary{shapes.geometric, arrows}
\usetikzlibrary{positioning}

\tikzset{
    startstop/.style={rectangle, rounded corners, minimum width=2.6cm,
        minimum height=1cm, text centered, draw=black, fill=gray!20},
    process/.style={rectangle, minimum width=3.2cm, minimum height=1cm,
        text centered, draw=black, fill=blue!20},
    decision/.style={diamond, aspect=2, text centered,
        draw=black, fill=green!20, inner sep=1pt},
    arrow/.style={thick,->,>=stealth}
}
\shorttitle{Magnetar Origin of GRB Plateaus}
\shortauthors{Dong et al.}

\begin{document}

\title{Diverse Morphologies of GRB X-Ray Plateaus within a Common Magnetar Framework}

\correspondingauthor{Yong-Feng Huang}
\email{hyf@nju.edu.cn}

\author[0009-0000-0467-0050]{Xiao-Fei Dong}
\affiliation{School of Astronomy and Space Science, Nanjing
University, Nanjing 210023, China}
\author[0000-0001-7199-2906]{Yong-Feng Huang}
\affiliation{School of Astronomy and Space Science, Nanjing
University, Nanjing 210023, China}
\affiliation{Key Laboratory of
Modern Astronomy and Astrophysics (Nanjing University), Ministry
of Education, China}
\author[0000-0001-9227-3716]{Nurimangul Nurmamat}
\affiliation{Guangxi Key Laboratory for Relativistic Astrophysics,
School of Physical Science and Technology, Guangxi University,
Nanning 530004, China}
\author[0000-0002-2191-7286]{Chen Deng}
\affiliation{School of Astronomy and Space Science, Nanjing
University, Nanjing 210023, China}
\affiliation{Key Laboratory of
Modern Astronomy and Astrophysics (Nanjing University), Ministry
of Education, China}
\author[0000-0002-6189-8307]{Ze-Cheng Zou}
\affiliation{School of Astronomy and Space Science, Nanjing
University, Nanjing 210023, China}
\author[0000-0001-7943-4685]{Fan Xu}
\affiliation{Institute of Space Weather, Nanjing University of
Information Science and Technology, Nanjing 210023, China}
\author[0000-0002-2162-0378]{Abdusattar Kurban}
\affiliation{State Key Laboratory of Radio Astronomy and
Technology, Xinjiang Astronomical Observatory, CAS, 150 Science
1-Street, Urumqi, Xinjiang 830011, China}
\affiliation{Xinjiang
Key Laboratory of Radio Astrophysics, Urumqi, Xinjiang 830011,
China}
\author[0009-0002-8460-1649]{Chen Du}
\affiliation{School of Astronomy and Space Science, Nanjing
University, Nanjing 210023, China}
\author[0000-0002-5238-8997]{Chen-Ran Hu}
\affiliation{School of Astronomy and Space Science, Nanjing
University, Nanjing 210023, China}
\author[0000-0001-9648-7295]{Jin-Jun Geng}
\affiliation{Purple Mountain Observatory, Chinese Academy of
Sciences, Nanjing 210023, China}

\begin{abstract}

The origin of the X-ray plateau phase in gamma-ray bursts (GRBs)
remains an open problem. In particular, it is unclear whether GRBs
with different temporal morphologies (i.e., with a rising, flat, or
decaying plateau) arise from a common underlying mechanism.
Although magnetar energy injection is a leading explanation,
previous studies have primarily inferred magnetar properties on a
burst-by-burst basis and have not tested the model at the
population level. Here we perform the first hierarchical
population inference of magnetar parameters for a uniform sample
of 185 long GRBs with X-ray plateaus within a conditional Poisson
point-process framework. It is found that the observed plateau
population is well reproduced by physically plausible magnetar
populations. The inferred parameter distributions show no strong
statistical separation among subclasses with different plateau
morphologies. Nevertheless, all subclasses show a substantial
intrinsic luminosity scatter, $\sigma_{L,\rm int}\sim0.5$--1.0 dex,
whereas the intrinsic duration scatter remains considerably
smaller. The results provide a population-level test of the
magnetar interpretation of GRB X-ray plateaus, showing that the
observed diversity of plateau morphologies does not require
distinct magnetar populations.

\end{abstract}


\section{Introduction} \label{sec:intro}
The X-ray plateau is one of the most intriguing features observed
in the early afterglows of gamma-ray bursts (GRBs)
\citep{2006Natur.444.1044G, 2006MNRAS.366L..13G,
2006ApJ...642..354Z}. Its temporal behavior spans a wide range,
including rising, nearly flat, and shallow-decaying plateaus
\citep{2019ApJS..245....1T, 2023ApJ...943..126D,
2023A&A...675A.117R, 2024ApJ...960...77D}. Analyses of X-ray afterglows,
prompt-emission properties, and population statistics have
consistently shown that GRBs exhibiting different plateau
morphologies share broadly similar observational characteristics
\citep{2020ApJ...903...18S, 2024ApJ...960...77D,
2026ApJ..1003..227D}. Although numerous models have been proposed
to explain these behaviors \citep{2006MNRAS.369..197F,
2007ApJ...670..565L, 2008MNRAS.391L..79D, 2009ApJ...690L.118Y}, it
remains unclear whether the observed diversity reflects
intrinsically distinct populations or a common mechanism operating
under different conditions \citep{2023MNRAS.525.5204B,
2024A&A...692A..73G, 2025JHEAp..4700384L}.

Among the proposed explanations for GRB X-ray plateaus,
energy injection from a newborn millisecond magnetar is one of
the leading scenarios
\citep{1998A&A...333L..87D, 1998PhRvL..81.4301D,
2001ApJ...552L..35Z, 2006MNRAS.369..197F}. As the magnetar
spins down, its rotational energy is gradually transferred to
the external blast wave, naturally producing a plateau phase
in the afterglow emission. In the standard dipole spin-down
model, the plateau luminosity and characteristic timescale are
primarily controlled by the surface magnetic field strength
$B_{\rm p}$ and initial spin period $P_0$, allowing plateau
observations to probe the properties of the central engine
\citep{2011MNRAS.413.2031M, 2013MNRAS.431.1745G,
2021ApJ...922..102H, 2024ApJ...974..133T}.

Previous studies of GRB X-ray plateaus within the magnetar
framework have largely relied on event-by-event analyses,
either through analytic inversion of the dipole spin-down
relations \citep{2010ApJ...715..477Y, 2014MNRAS.443.1779R, 2014ApJ...785...74L}
or direct fitting of plateau light curves
\citep{2018ApJ...869..155S, 2026ApJ..1000...97D}.
Early applications used a small number of X-ray plateaus to
constrain the magnetar spin period and magnetic field strength
\citep{2010MNRAS.402..705L}, and were subsequently extended to
short GRBs and extended-emission bursts
\citep{2013MNRAS.430.1061R, 2014MNRAS.438..240G, 2015ApJ...805...89L}.
More recent studies have applied these methods to larger GRB
samples to investigate the distributions and correlations of
the inferred magnetar parameters
\citep{2025ApJS..280...45L, 2026ApJ..1004L..11Z, 2026arXiv260718698L}.

In this work, we infer the intrinsic magnetar population
using a Poisson point-process framework. It treats the observed
events as a realization of an intrinsic source population filtered
by observational selection, thereby enabling inference on the
underlying population rather than the detected sample alone.
Early applications of the approach to GRB population studies have been
carried out by \citet{1995ApJS...96..261L, 1998ApJ...502...75L,
1998ApJ...502..108L}. More recently, it has been used to infer
luminosity and prompt-emission distribution
\citep{2015MNRAS.451..126S, 2023A&A...680A..45S}, 
delay-time distribution
\citep{2026arXiv260103861P}, and jet geometry and angular
structure \citep{2020ApJ...895..108F, 2020ApJ...893...38B} of
GRBs. It is also widely used for gravitational-wave
\citep{2015PhRvD..91b3005F, 2018MNRAS.477.4685B,
2018ApJ...863L..41F, 2019MNRAS.486.1086M}, fast radio burst
\citep{2026arXiv260709109W}, and exoplanet
\citep{2014ApJ...795...64F} population studies.

The paper is organized as follows.
Section~\ref{sec:method} describes the hierarchical Bayesian framework.
Section~\ref{sec:results} presents the inferred magnetar populations.
Section~\ref{sec:test} examines the robustness of the population inference. 
Section~\ref{sec:conclusion} summarizes the main conclusions.
Section~\ref{sec:Discussion} discusses the implications of our findings.

\section{METHODS}
\label{sec:method}

We infer the intrinsic magnetar population by
forward modeling the connection between the central-engine parameters
and the observed X-ray plateau sample.
The analysis proceeds in five steps. 

First, we specify the intrinsic distributions of the
magnetar spin period and magnetic field.
These parameters are mapped to the plateau luminosity and
spin-down timescale through the magnetic-dipole model,
including intrinsic scatter. Redshifts are then drawn from
the assumed distribution to derive the fluxes and
observer-frame break times, which determine the selection probability.
The resulting selection-modified population enters the
Poisson point-process likelihood to constrain the magnetar population.
Finally, the required integrals are evaluated by Monte Carlo forward modeling.

\subsection{Intrinsic magnetar population}

We describe the intrinsic magnetar population using the
initial spin period $P$ and dipole magnetic field strength $B$.
Previous event-by-event and population studies indicate that
the inferred magnetar parameters occupy a relatively restricted region
of the $P$--$B$ plane
\citep{2014MNRAS.443.1779R, 2015ApJ...805...89L,
2018ApJ...869..155S, 2018ApJS..236...26L, 2026ApJ..1000...97D},
and recent population analyses suggest that their distributions are
approximately log-normal \citep{2025ApJS..280...45L,
2026ApJ..1004L..11Z}.

We therefore assume
\begin{equation}
\log_{10}(P/{\rm ms}) \sim \mathcal N(P_{\rm c},\sigma_P^2),
\qquad
\log_{10}(B/{\rm G}) \sim \mathcal N(B_{\rm c},\sigma_B^2),
\end{equation}
where $P_{\rm c}$ and $B_{\rm c}$ are the population means in
logarithmic space, and $\sigma_P$ and $\sigma_B$ are the
corresponding intrinsic dispersions.
For simplicity, $P$ and $B$ are assumed to be independent, so that
their joint population distribution is
\begin{equation}
p(P,B\mid\phi_{\rm pop}) = p(P\mid P_{\rm c},\sigma_P)\,
p(B\mid B_{\rm c},\sigma_B),
\end{equation}
with $\phi_{\rm pop} = (P_{\rm c},\sigma_P,B_{\rm c},\sigma_B)$.

\subsection{Forward model from magnetar parameters to plateau observables}

For a given pair \((P,B)\), the magnetic-dipole spin-down
model predicts a characteristic luminosity \(L_0\) and rest-frame
spin-down timescale \(T_0\). Adopting the neutron star radius as
\(R_{\rm NS}=10^6\,{\rm cm}\) and moment of inertia as
\(I=10^{45}\,{\rm g\,cm^2}\), we have \citep{2001ApJ...552L..35Z,
2010MNRAS.402..705L}
\begin{equation}
L_0(P,B) = 10^{49} \left(\frac{B}{10^{15}\,{\rm G}}\right)^2 \left(\frac{P}{1\,{\rm ms}}\right)^{-4} \,{\rm erg\,s^{-1}},
\end{equation} and
\begin{equation}
T_0(P,B) = 10^{3} \left(\frac{B}{10^{15}\,{\rm G}}\right)^{-2} \left(\frac{P}{1\,{\rm ms}}\right)^2 \,{\rm s}.
\end{equation}

We do not introduce the X-ray radiative efficiency as an
explicit free parameter. Instead, deviations from the
idealized spin-down scalings are modeled phenomenologically
with log-normal intrinsic scatter:
\begin{equation}
\log_{10}L
=
\log_{10}L_0(P,B)+\epsilon_L,
\qquad
\epsilon_L\sim \mathcal N(0,\sigma_{L,\rm int}^2),
\end{equation}
\begin{equation}
\log_{10}T
=
\log_{10}T_0(P,B)+\epsilon_T,
\qquad
\epsilon_T\sim \mathcal N(0,\sigma_{T,\rm int}^2).
\end{equation}
Here \(\sigma_{L,\rm int}\) and \(\sigma_{T,\rm int}\) describe
the population-level dispersion around the idealized magnetar
relations. They absorb unmodeled diversity in radiative
efficiency, neutron-star structure, magnetic-field geometry, and
outflow properties \citep{2014MNRAS.443.1779R,
2023ApJ...949L..32D}. The residuals $\epsilon_L$ and
$\epsilon_T$ are assumed to be mutually independent, and
independent of $P$ and $B$ as well.

The forward model therefore defines the mapping
$(P,B)\longrightarrow(L,T)$.
Together with the intrinsic-scatter parameters, the full set of
population-shape parameters is $\phi= (P_{\rm c},\sigma_P,B_{\rm c},
\sigma_B, \sigma_{L,\rm int},\sigma_{T,\rm int})$.

\subsection{Redshift distribution and selection function}

Redshift is treated as a population-level latent variable
and is marginalized over rather than included as an explicit
event-level variable in the likelihood. It enters the analysis
through the observer-frame transformation and catalog selection.

For the fiducial model, we assume that the long-GRB
occurrence rate follows the cosmic star-formation history
\citep{2006MNRAS.372.1034D, 2008ApJ...683L...5Y, 2025ApJ...990...69K,
2025ApJ...994..185B}.
The normalized observer-frame redshift density is
\begin{equation}
p(z)= \frac{ \dot{\rho}(z)\,[dV(z)/dz]/(1+z) }
{ \int_{z_{\rm min}}^{z_{\rm max}} \dot{\rho}(z)\,[dV(z)/dz]/(1+z)\,dz },
\end{equation}
where the factor \((1+z)^{-1}\)
accounts for cosmological time dilation.
The differential comoving volume is
\begin{equation}
\frac{dV(z)}{dz} = \frac{c}{H_0} \frac{4\pi D_{\rm L}^2(z)}{(1+z)^2}
\frac{1}{ \sqrt{\Omega_{\rm M}(1+z)^3+\Omega_\Lambda} }.
\end{equation}
We adopt a flat \(\Lambda\)CDM cosmology with
\(H_0=67.3~{\rm km~s^{-1}~Mpc^{-1}}\),
\(\Omega_{\rm M}=0.315\), and \(\Omega_\Lambda=1-\Omega_{\rm M}\)
\citep{2014A&A...571A..16P, 2020A&A...641A...6P}.
The star-formation-rate density is parameterized as
\citep{2006ApJ...651..142H}
\begin{equation}
\dot{\rho}(z)\propto \frac{a_{\rm sfr}+b_{\rm sfr}z}
{1+(z/c_{\rm sfr})^{d_{\rm sfr}}},
\end{equation}
with \((a_{\rm sfr},b_{\rm sfr},c_{\rm sfr},d_{\rm sfr}) =(0.014,0.140,2.98,4.55)\) \citep{2022MNRAS.513.1078D}.

For each simulated event, a redshift is drawn from $p(z)$.
The corresponding flux and observer-frame break time are then
calculated as
\begin{equation}
F=\frac{L}{4\pi D_L^2(z)},
\qquad
t_b=T(1+z).
\end{equation}
These quantities determine the probability that the event enters
the observed plateau sample.

The observed sample is not an unbiased realization of the
intrinsic population. Identifying an X-ray plateau requires both
sufficient flux and a measurable break within the available
temporal coverage \citep{2014ApJ...783...24L,
2015ApJ...800...31D}. We model these two dominant effects with a
separable selection function,
\begin{equation}
S(L,t_b,z)=S_F(F)\,S_T(t_b).
\end{equation}
Here the flux-selection term is modeled with a smooth logistic
function,
\begin{equation}
S_F(F) = \left[ 1+\exp\left( -\frac{\log_{10}(F/F_{\rm th})}{\Delta_F} \right) \right]^{-1},
\end{equation}
where \(F_{\rm th}\) is the effective flux threshold
and \(\Delta_F\) controls the transition width.
The duration-selection term is
\begin{equation}
S_T(t_b) = \left[ 1+\exp\left( -\frac{\log_{10}(t_b/t_{\rm min})}
{\Delta_T} \right) \right]^{-1}
\left[ 1+\exp\left( \frac{\log_{10}(t_b/t_{\rm max})}{\Delta_T} \right) \right]^{-1}.
\end{equation}
We adopt $\Delta_F=\Delta_T=0.5$, corresponding to a smooth
transition at the nominal selection boundaries.

\subsection{Conditional Poisson point-process likelihood}

We now connect the forward model to the observed catalog.
For the $i$th GRB, we denote the measured plateau quantities by
\begin{equation}
d_i=(\ell_i,\tau_i),
\end{equation}
where \(\ell_i=\log_{10}L_i\) and \(\tau_i=\log_{10}t_{b,i}\).
The corresponding measurement uncertainties are $\sigma_{\ell,i}$
and $\sigma_{\tau,i}$.
For a modeled GRB location \(y=(\ell,\tau)\),
where $\ell=\log_{10}L$ and $\tau=\log_{10}t_b$,
the measurement probability is described by \citep{2007ApJ...665.1489K}
\begin{equation}
K_i(d_i\mid y)
=
\mathcal N(\ell_i\mid \ell,\sigma_{\ell,i}^2)
\mathcal N(\tau_i\mid \tau,\sigma_{\tau,i}^2),
\end{equation}
where \(\mathcal N(a\mid b,\sigma^2)\) denotes a Gaussian density
evaluated at \(a\), with mean \(b\) and variance \(\sigma^2\).
This kernel gives the probability of observing $d_i$ if the modeled plateau location of the event is $y$.

For a given population model \(\phi\),
the selected intensity density at the location of event $i$ is
\begin{equation}
\lambda(d_i\mid\phi) =
\int dP\,dB\, p(P,B\mid\phi) \int dz\,p(z) \int dy\,
p(y\mid P,B,z,\phi)\, S(y,z)\, K_i(d_i\mid y).
\label{eq:event_intensity}
\end{equation}
Here $p(P,B\mid\phi)$ describes the intrinsic magnetar population,
$p(z)$ is the adopted redshift distribution,
and $p(y\mid P,B,z,\phi)$ represents the stochastic forward model.
The corresponding population-averaged detection probability is
\begin{equation}
\omega(\phi) = \int dP\,dB\, p(P,B\mid\phi) \int dz\,p(z) \int dy\,
p(y\mid P,B,z,\phi)\, S(y,z).
\label{eq:omega}
\end{equation}

Let \(R\) denote the expected number of intrinsic events
over the survey exposure before selection.
Then the number of detected events is
$N_{\rm obs}(\phi,R)=R\,\omega(\phi)$.
The full Poisson point-process likelihood is \citep{1995ApJS...96..261L, 1998ApJ...502...75L, 1998ApJ...502..108L, 2019MNRAS.486.1086M, 2019PASA...36...10T}
\begin{equation}
\mathcal L(\phi,R)
=
\exp[-R\omega(\phi)]
\prod_{i=1}^{N_{\rm obs}}
R\,\lambda(d_i\mid\phi).
\label{eq:full_ppp}
\end{equation}

The absolute completeness of the plateau sample is not
sufficiently well characterized to constrain the plateau-event
rate. We therefore condition on the observed number of events,
\(N_{\rm obs}\), and use the likelihood only to constrain the
population shape \citep{2018ApJ...863L..41F, 2019MNRAS.486.1086M}.
The resulting conditional Poisson point-process (CPP) likelihood
is
\begin{equation}
\mathcal L_{\rm CPP}(\phi)
\propto
\prod_{i=1}^{N_{\rm obs}}
\frac{\lambda(d_i\mid\phi)}
{\omega(\phi)},
\label{eq:cpp_likelihood}
\end{equation}
or equivalently
\begin{equation}
\ln\mathcal L_{\rm CPP}(\phi)
=
\sum_{i=1}^{N_{\rm obs}}
\ln\lambda(d_i\mid\phi)
-
N_{\rm obs}\ln\omega(\phi)
+
{\rm const}.
\label{eq:log_cpp_likelihood}
\end{equation}
Here the first term measures agreement between the predicted
population and the observed events, while the second corrects for
the population-averaged selection probability. The same expression
is obtained by marginalizing the full Poisson point-process
likelihood over $R$ using the scale-invariant prior $\pi(R)\propto
1/R$ \citep{2018ApJ...863L..41F, 2019MNRAS.486.1086M}.

\subsection{Monte Carlo evaluation and posterior sampling}

The integrals entering $\lambda(d_i\mid\phi)$ and
$\omega(\phi)$ are evaluated by Monte Carlo forward modeling,
as the convolution of the latent population, stochastic
forward model, measurement kernel, and selection function does
not lead to closed-form expressions.
For each trial hyperparameter set $\phi$,
we generate $N_{\rm sim}$ events by drawing
\begin{equation}
(P_j,B_j)\sim p(P,B\mid\phi),
\qquad
z_j\sim p(z),
\qquad
j=1,\ldots,N_{\rm sim}.
\end{equation}
Intrinsic luminosity and duration scatter are then applied, and
the events are transformed into $y_j=(\ell_j,\tau_j)$ . A
selection weight $w_j=S(y_j,z_j)$ is assigned to each draw.

The population-averaged detection probability is estimated as
\begin{equation}
\widehat{\omega}(\phi)
=
\frac{1}{N_{\rm sim}}
\sum_{j=1}^{N_{\rm sim}}
w_j.
\end{equation}
For each observed GRB \(d_i=(\ell_i,\tau_i)\), the selected intensity
density is estimated by
\begin{equation}
\widehat{\lambda}(d_i\mid\phi)
=
\frac{1}{N_{\rm sim}}
\sum_{j=1}^{N_{\rm sim}}
w_j\,K_i(d_i\mid y_j).
\end{equation}
For robustness, the observational-error kernel is evaluated using
effective uncertainties
\begin{equation}
s_{\ell,i} = \max(\sigma_{\ell,i},0.05),
\qquad
s_{\tau,i} = \max(\sigma_{\tau,i},0.05),
\end{equation}
so that
\begin{equation}
K_i(d_i\mid y_j)
=
\mathcal N(\ell_i\mid\ell_j,s_{\ell,i}^2)
\mathcal N(\tau_i\mid\tau_j,s_{\tau,i}^2).
\label{KDE}
\end{equation}
The CPP likelihood is therefore approximated as
\begin{equation}
\ln \widehat{\mathcal L}_{\rm CPP}(\phi)
=
\sum_{i=1}^{N_{\rm obs}}
\ln \widehat{\lambda}(d_i\mid\phi)
-
N_{\rm obs}\ln \widehat{\omega}(\phi).
\end{equation}
To reduce Monte Carlo noise, we use common random numbers by fixing
the random quantiles used to generate $P_j$, $B_j$, and $z_j$,
as well as the standard normal variates used for the intrinsic
luminosity and duration scatters.

The posterior distribution is
\begin{equation}
p(\phi\mid{d_i})
\propto
\widehat{\mathcal L}_{\rm CPP}(\phi)\ p(\phi).
\end{equation}
We sample the posterior distribution using the \texttt{emcee}
Markov chain Monte Carlo sampler \citep{2013PASP..125..306F},
using 48 walkers and 30000 steps per walker.
The first 10000 steps are discarded as burn-in.
Independent uniform priors over broad physically motivated ranges are
adopted for all model parameters, as summarized in
Table~\ref{tab:cpp_params}.

\section{Results} \label{sec:results}

The uniform sample of 185 long GRBs with well-defined
X-ray plateaus compiled by \cite{2026ApJ..1003..227D} are analyzed
in our study. All GRBs in this sample have measured redshifts. We
adopt the observed break time $T_{a,\rm obs}$ and the K-corrected
plateau luminosity $L_{\rm X}$, where $L_{\rm X}$ was derived
using the corresponding redshift, as inputs to the CPP analysis.

The plateau morphology is characterized by the
temporal slope $\alpha_1$ of the plateau segment.
Following the classification scheme of \cite{2026ApJ..1003..227D},
the 185 GRBs are divided into rising ($-0.5<\alpha_1\leq-0.1$, 16 GRBs),
flat ($-0.1<\alpha_1\leq0.1$, 75 GRBs),
and decaying ($0.1<\alpha_1\leq0.5$, 94 GRBs) subclasses.
The boundaries of $\pm0.1$ are phenomenological
rather than theoretically predicted critical values.
They are adopted to distinguish nearly flat plateaus from
those showing a clear rising or decaying trend and to test
whether different plateau morphologies correspond to distinct populations.

\subsection{CPP Analysis of the Full Plateau GRB Sample}

Following the CPP framework described in
Section~\ref{sec:method}, we analyze the full sample
of 185 plateau GRBs using the K-corrected plateau
luminosities \(L_{\rm X}\) and observed break times
\(T_{a,\rm obs}\).
The minimum observed plateau flux is used as an empirical flux scale,
$\log_{10}({F_{\rm th}}/{\rm 1\ erg~cm^{-2}~s^{-1}})=-12.76$.
We adopt
$\log_{10}({t_{\rm min}}/{\rm 1\ s})=1$ and
$\log_{10}({t_{\rm max}}/{\rm 1\ s})=6$,
consistent with the observed range of plateau measurements
\citep{2012ApJ...758...27L, 2019ApJS..245....1T}.
The adopted priors and posterior constraints
on the population parameters are summarized in
Table~\ref{tab:cpp_params} and also shown in Figure~A2.

\begin{figure*}[ht!]
\centering
\includegraphics[width=0.95\textwidth, trim=0 0 0 0, clip]{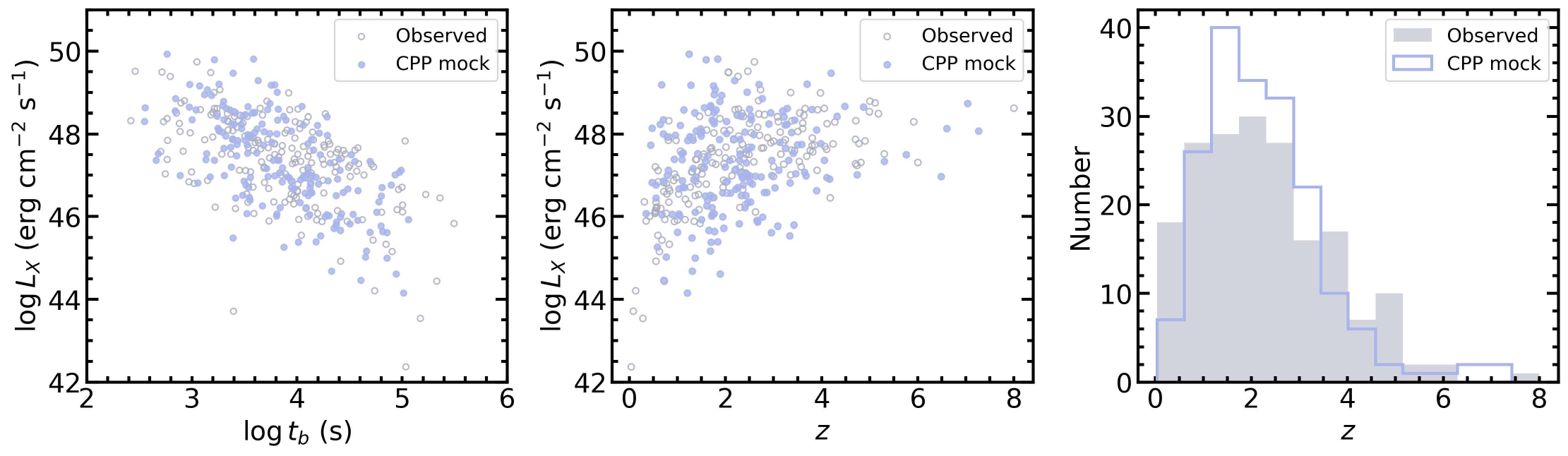}
\caption{Parameter distributions of the observed sample and the
CPP mock bursts. The three panels show the distributions in the
$L_{X}-t_{b}$ plane, the $L_{X}-z$ plane, and the redshift
distribution. The simulated source sample includes all 185 plateau
events used in this work.} \label{fig1}
\end{figure*}

Figure~\ref{fig1} shows a simulated GRB population generated from
the posterior-median parameter values, in the $L_{\rm X}$--$t_{\rm b}$
plane, the $L_{\rm X}$--$z$ plane, and the redshift distribution.
The simulated
population provides an adequate description of the observed
marginal distributions. To quantify the agreement, we performed
Kolmogorov--Smirnov (KS) tests on the luminosity, break-time,
and redshift distributions. The resulting $p$-values are
$p_{L_{\rm X}}=0.106$, $p_{t_{\rm b}}=0.560$, and $p_z=0.217$,
respectively.
None of these tests rejects consistency between
the simulated and observed distributions at the 5\% significance
level. The redshift comparison should be interpreted as a
consistency check of the adopted SFR-based redshift model after
selection, rather than as an independent inference of the
intrinsic redshift distribution.

\begin{figure*}[ht!]
\centering
\includegraphics[width=0.55\textwidth, trim=0 0 0 0, clip]{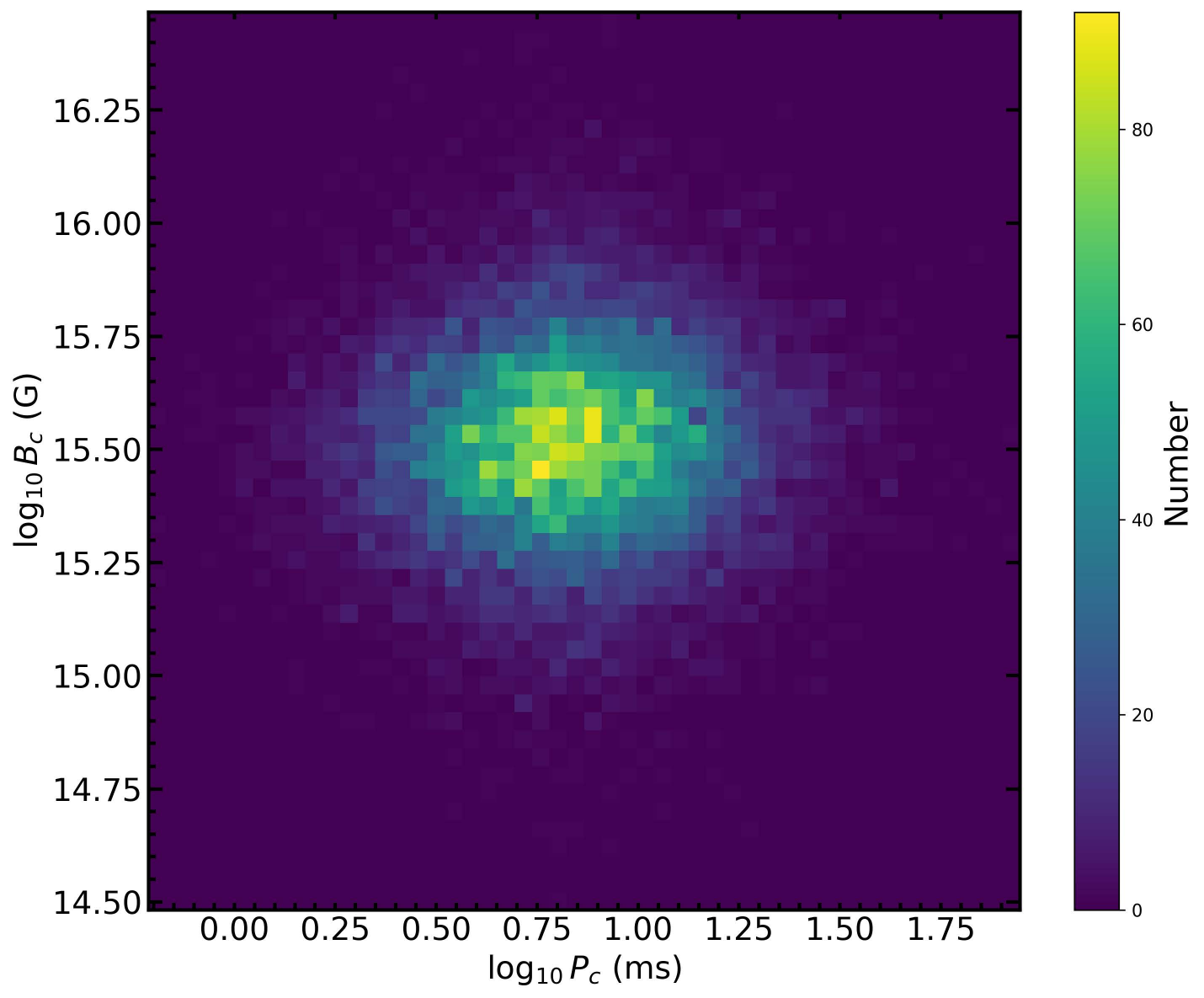}
\caption{Two-dimensional distribution of magnetar spin periods
and magnetic fields for the full plateau sample, reconstructed from 15,000 Monte Carlo draws
from the posterior population. The color scale indicates the
number of samples in each bin. } \label{fig2}
\end{figure*}

Figure~\ref{fig2} presents the reconstructed posterior
distribution in the $\log_{10}B_{\rm c}$--$\log_{10}P_{\rm c}$ plane.
The distribution map is obtained through Monte Carlo sampling of the
latent magnetar population implied by the posterior parameter
distributions. A well-defined concentration region is evident,
demonstrating that the observed plateau population provides
meaningful constraints on the characteristic magnetar properties.
The posterior peaks around $\log_{10}P_{\rm c}\simeq0.83$
and $\log_{10}B_{\rm c}\simeq15.52$, corresponding to
spin periods of a few milliseconds and surface magnetic fields of
order $10^{15}\mathrm{G}$.

These values are broadly consistent with the expected properties
of magnetars proposed as the central engines of GRB X-ray plateaus
\citep{2025ApJS..280...45L, 2026ApJ..1000...97D, 2026ApJ..1004L..11Z}.
Previous event-by-event works generally inferred millisecond initial
spin periods and surface dipole magnetic fields of order $10^{14}-10^{16}$ G,
broadly overlapping with the characteristic ranges inferred here.

\begin{figure*}[ht!]
\centering
\includegraphics[width=0.95\textwidth, trim=0 0 0 0, clip]{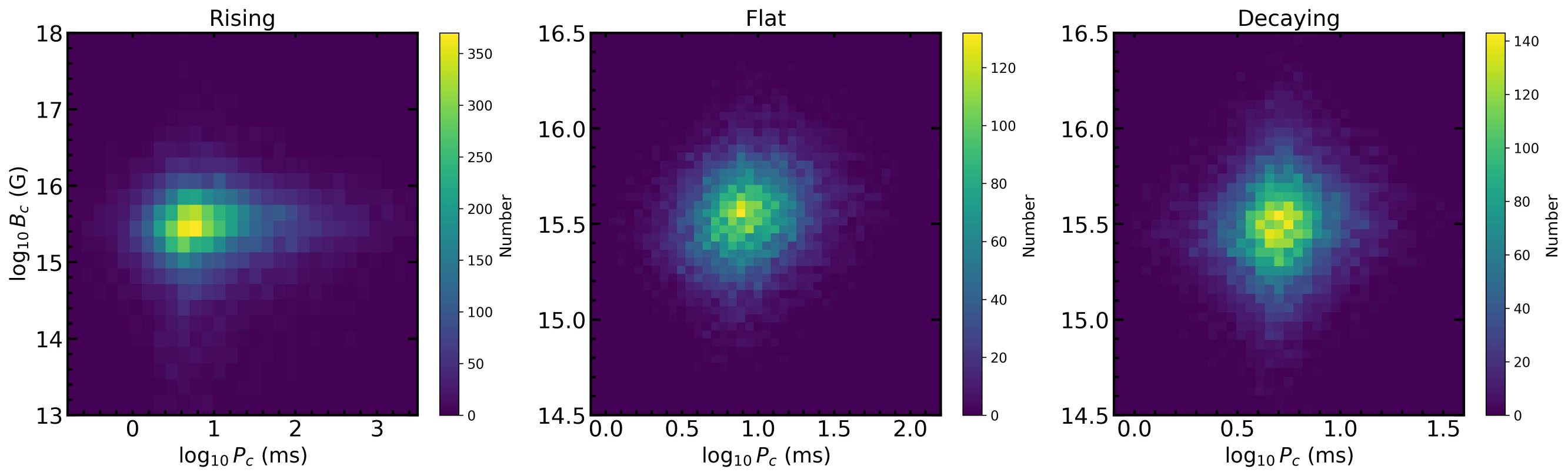}
\caption{Two-dimensional posterior distribution of the three subclasses in the
$\log_{10}B_{\rm c}-\log_{10}P_{\rm c}$ plane. The color scale
indicates the number of posterior samples in each bin. From left
to right, the panels correspond to the rising, flat, and decaying
samples, respectively. Each panel is constructed from 15,000
posterior samples for the corresponding plateau class. }
\label{fig3}
\end{figure*}

\subsection{CPP Analysis of the Rising, Flat, and Decaying Subsamples}

To investigate whether different plateau morphologies require
distinct magnetar populations, we apply the CPP framework
separately to the rising, flat, and decaying subsamples. The
resulting parameter constraints are summarized in
Table~\ref{tab:cpp_params}, and the corresponding posterior
distributions are shown in Figure~A2.
The inferred magnetar properties are broadly consistent among the
three subclasses. The characteristic magnetic fields cluster around
$\log_{10}B_{\rm c}\sim15.5$, while the characteristic spin
periods lie within the range $\log_{10}P_{\rm c}\sim0.7-0.9$.

Figure~\ref{fig3} reconstructs the posterior distributions
in the $\log_{10}B_{\rm c}-\log_{10}P_{\rm c}$ plane.
The high-probability regions exhibit substantial overlap, and no
obvious abnormal distribution is evident among the three subclasses.
Among the three subclasses, the rising sample exhibits the broadest
posterior distribution. This is probably due to its substantially
smaller sample size $(N=16)$, which provides weaker statistical
constraints than the flat and decaying samples. Nevertheless, its
posterior region remains similar with those inferred from
the other two subclasses.

\begin{table*}[ht]
\centering
\caption{CPP model parameters and posterior constraints.}
\label{tab:cpp_params}
\normalsize
\setlength{\tabcolsep}{10pt}

\begin{tabular}{lccccc}
\hline
Parameter & Prior range & Total (N=185)& Rising (N=16) & Flat (N=75) & Decaying (N=94)\\
\hline

$P_{\rm c}$ (log$_{10}$ ms)
& $[-1,\,2]$
& $0.83_{-0.04}^{+0.04}$
& $0.90_{-0.28}^{+0.62}$
& $0.92_{-0.11}^{+0.11}$
& $0.72_{-0.06}^{+0.07}$ \\

$\sigma_P$ (dex)
& $[0.05,\,2]$
& $0.27_{-0.03}^{+0.02}$
& $0.46_{-0.21}^{+0.30}$
& $0.25_{-0.07}^{+0.05}$
& $0.19_{-0.10}^{+0.07}$ \\

$B_{\rm c}$ (log$_{10}$ G)
& $[13,\,17]$
& $15.52_{-0.05}^{+0.05}$
& $15.46_{-0.24}^{+0.19}$
& $15.54_{-0.09}^{+0.10}$
& $15.49_{-0.07}^{+0.07}$ \\

$\sigma_B$ (dex)
& $[0.1,\,2]$
& $0.19_{-0.06}^{+0.06}$
& $0.34_{-0.16}^{+0.28}$
& $0.18_{-0.05}^{+0.06}$
& $0.24_{-0.10}^{+0.07}$ \\

$\sigma_{L,\mathrm{int}}$ (dex)
& $[0.01,\,1.5]$
& $0.87_{-0.09}^{+0.09}$
& $0.46_{-0.29}^{+0.36}$
& $1.06_{-0.16}^{+0.15}$
& $0.71_{-0.18}^{+0.14}$ \\

$\sigma_{T,\mathrm{int}}$ (dex)
& $[0.01,\,1.5]$
& $0.12_{-0.07}^{+0.15}$
& $0.24_{-0.16}^{+0.21}$
& $0.17_{-0.09}^{+0.12}$
& $0.23_{-0.16}^{+0.16}$ \\

\hline
\end{tabular}

\vspace{2mm}

\begin{minipage}{0.92\textwidth}
\small \textit{Note.} Uniform priors are adopted over the ranges
listed in the second column. The third through sixth columns
summarize the posterior constraints and the corresponding 68\%
confidence interval derived from the corresponding samples.
\end{minipage}
\end{table*}

\subsection{Comparison Among All the Plateau Samples}

\begin{figure*}[ht!]
\centering
\includegraphics[width=0.95\textwidth, trim=0 0 0 0, clip]{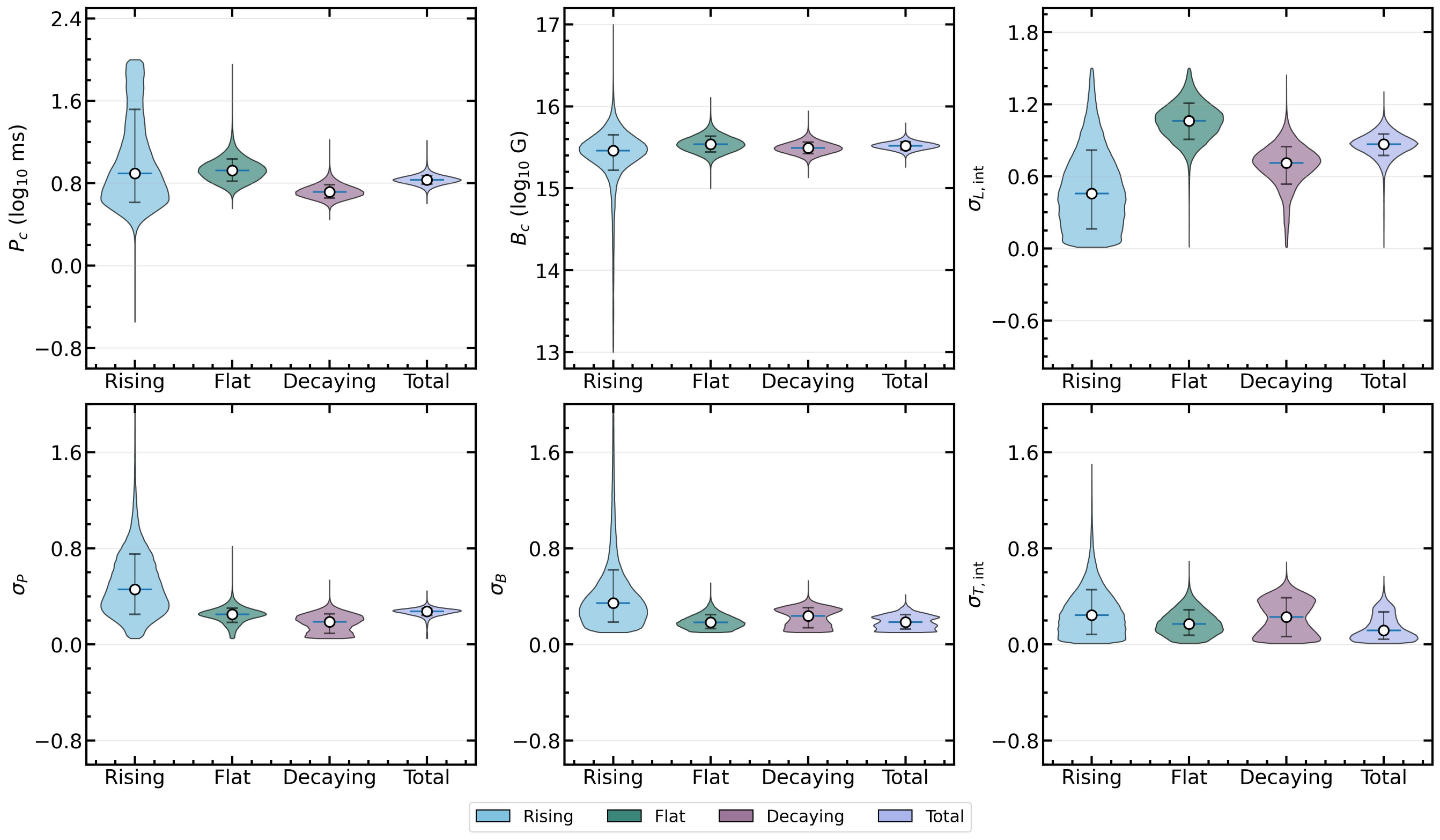}
\caption{ Violin plots of the posterior distributions of the six
CPP parameters inferred for the Rising, Flat, Decaying, and Total
plateau samples. The panels show $P_{\rm c}$, $\sigma_{P}$,
$B_{\rm c}$, $\sigma_{B}$, $\sigma_{L,\rm int}$, and $\sigma_{T,\rm
int}$, respectively. White markers and error bars denote the
median values and 68\% confidence intervals. The best-fit
parameter estimates are listed in Table~\ref{tab:cpp_params}. }
 \label{fig4}
\end{figure*}

\begin{figure*}[ht!]
\centering
\includegraphics[width=0.95\textwidth, trim=0 0 0 0, clip]{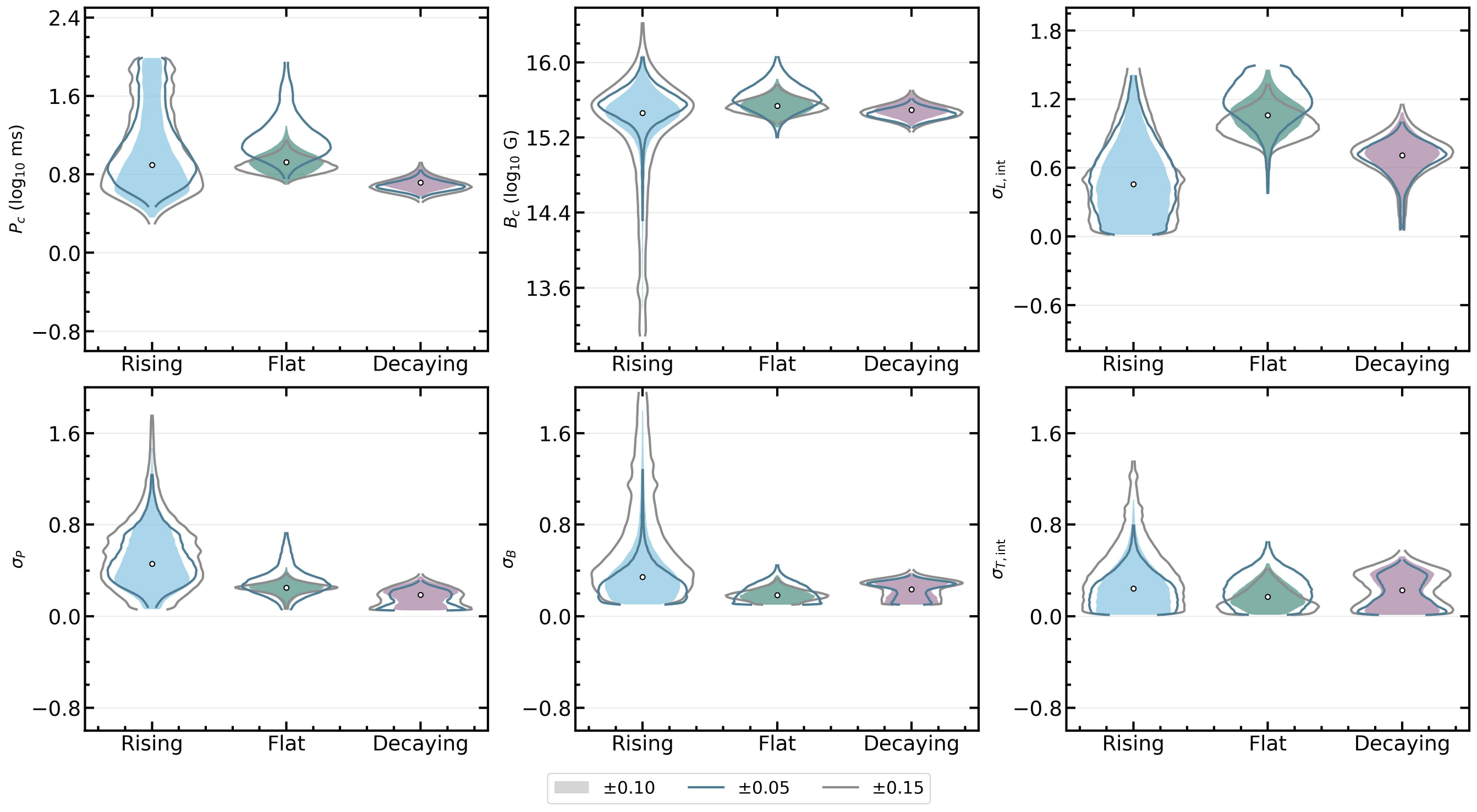}
\caption{
Violin plots showing the robustness test for
different plateau slope criteria. The filled violins correspond
to the fiducial boundary of $\pm0.10$ and are
identical to those shown in Figure~\ref{fig4}.
The blue violin outlines show the results for $\pm0.05$,
while the gray outlines correspond to $\pm0.15$.
Hollow circles denote the posterior medians of the fiducial results.}
\label{fig5}
\end{figure*}

Figure~\ref{fig4} presents violin plots of the posterior
distributions of the six CPP parameters inferred from the rising,
flat, decaying, and total samples. The violin widths represent
posterior probability densities, while the white markers and
error bars denote the median values and corresponding
68\% credible intervals.

The posterior distributions exhibit broad overlap among the full
sample and the three plateau subclasses, while the rising
subsample yields broader constraints owing to its smaller sample
size. To quantify these similarities, we computed the overlap
coefficient (OVL; \citealt{Inman01011989}), which ranges from 0 (no
overlap) to 1 (complete overlap), from pairwise comparisons of the
posterior distributions for the three subclasses. This analysis
excludes the total sample because it is not independent of the
subsamples. The magnetic-field parameters and intrinsic duration
scatter show moderate-to-strong overlaps (${\rm OVL}\sim0.5-0.8$),
whereas the spin-period centroid and intrinsic luminosity scatter
exhibit weaker overlaps (${\rm OVL}\sim0.2-0.5$). Overall, the
inferred parameter distributions do not exhibit strong statistical
separation.

A notable result is the relatively large intrinsic luminosity
scatter required by all samples,
$\sigma_{L,\rm int}\sim0.5$--1.0 dex,
which is systematically larger than the corresponding intrinsic
duration scatter. The flat sample exhibits the largest luminosity
dispersion, $\sigma_{L,\rm int}=1.06^{+0.15}_{-0.16}$ dex.
This indicates that, even within a common magnetar population,
plateau luminosities retain substantial source-to-source scatter
beyond measurement uncertainties and selection effects.

\section{Robustness Tests of the Population Inference}
\label{sec:test}

The three morphological subsamples defined using the fiducial boundaries at $\pm0.1$ yield broadly consistent magnetar population parameters. Since this boundary is phenomenological, we further test whether the inferred results depend on the adopted slope criterion.

To examine this effect, we repeat the classification
and the full CPP inference using alternative boundaries of
$\pm0.05$ and $\pm0.15$.
For $\pm0.05$, the Rising, Flat, and Decaying subsamples contain
20, 45, and 120 GRBs, respectively, while for $\pm0.15$ the
corresponding numbers are 13, 100, and 72.
The resulting posterior distributions are shown in Figure~\ref{fig5},
together with the fiducial $\pm0.1$ results.

As shown in Figure~\ref{fig5}, the posterior distributions obtained
with the alternative boundary criteria largely overlap with the
fiducial posterior ranges, indicating that our main conclusions
are robust to reasonable changes in the slope boundaries.
The Flat subsample shows the largest sensitivity to the boundary choice
because its membership changes at both the lower and upper classification boundaries,
with the most noticeable variation occurring in $P_{\rm c}$.
For the $\pm0.15$ criterion, the reduced sizes of the Rising
and Decaying subsamples result in broader posterior distributions,
but their inferred parameters remain generally consistent with the
fiducial results.
In particular, the magnetic-field parameters remain stable,
and $\sigma_{L,\rm int}$ remains larger than $\sigma_{T,\rm int}$
under all tested criteria.

We performed three additional robustness tests. First, we
replaced the fiducial star-formation-rate-based redshift
distribution with an empirical distribution derived from the
observed redshifts of the plateau sample. Second, we varied the
common width of the flux- and duration-selection functions,
$\Delta_F=\Delta_T$, from 0.3 to 0.7. Third, we increased the
minimum effective uncertainty in Equation~\ref{KDE} from 0.05 to
0.10 dex. In each test, the inferred magnetar population
parameters remained consistent with the fiducial results within
their credible intervals, and the main conclusions were
unchanged.

\section{Conclusion}
\label{sec:conclusion}

In this study, we applied a hierarchical Bayesian framework based
on a conditional Poisson point-process likelihood to infer the
population properties of magnetar central engines in GRB X-ray
plateaus. The method constrains the underlying magnetar population
within the adopted forward model while accounting for
observational selection effects and measurement uncertainties.
Applying the framework to a uniform sample of 185
\textit{Swift}/XRT plateau GRBs, we found that the posterior
distributions satisfactorily reproduce the observed distributions
of plateau luminosity, break time, and redshift.
The posterior distributions favor a physically plausible
population of highly magnetized millisecond magnetars,
with a characteristic spin period of $\sim6.8$ ms and
a magnetic field strength of $\sim3.3\times10^{15}$ G.

Applying the same framework to the rising, flat, and decaying
plateau subsamples, we find no strong statistical evidence for
distinct magnetar populations. The inferred parameter
distributions exhibit substantial overlap and remain broadly
consistent with those obtained for the full sample. Despite
this overall similarity, all subclasses require a relatively
large intrinsic luminosity scatter,
$\sigma_{L,\rm int}\sim0.5$--1.0 dex, whereas the intrinsic
duration scatter is considerably smaller. These results suggest
that different plateau morphologies can be understood within a
common magnetar framework, although additional sources of
luminosity diversity may still be present within the current
model.

\section{Discussion}
\label{sec:Discussion}
\subsection{Consistency with Traditional Magnetar Estimates}

To compare the Bayesian population inference with the
traditional event-by-event approach, we derive magnetar
parameters from the observed plateau luminosity and break time.
Following \cite{2001ApJ...552L..35Z} and
\citet{2015ApJ...805...89L}, we assume
$T=t_{\rm b}/(1+z)$ and $L_{\rm b}\simeq \eta L_{0}$,
where $\eta$ denotes the radiation efficiency.
The corresponding magnetic field and initial spin period are
\begin{equation}
B_{\rm p,15} = 2.05\, I_{45} R_{6}^{-3}
L_{0,49}^{-1/2} T_{3}^{-1},
\qquad
P_{0,-3} = 1.42\, I_{45}^{1/2}
L_{0,49}^{-1/2} T_{3}^{-1/2},
\end{equation}
where $B_{\rm p,15}=B_{\rm p}/10^{15}\ {\rm G}$,
$P_{0,-3}=P_{0}/10^{-3}\ {\rm s}$,
$L_{0,49}=L_{0}/10^{49}\ {\rm erg\ s^{-1}}$, and
$T_{3}=T/10^{3}\ {\rm s}$.
For consistency, we adopt the same neutron-star radius and
moment of inertia as assumed in the Bayesian population model.

\begin{figure*}[ht!] \centering
\includegraphics[width=0.6\textwidth, trim=0 0 0 0, clip]{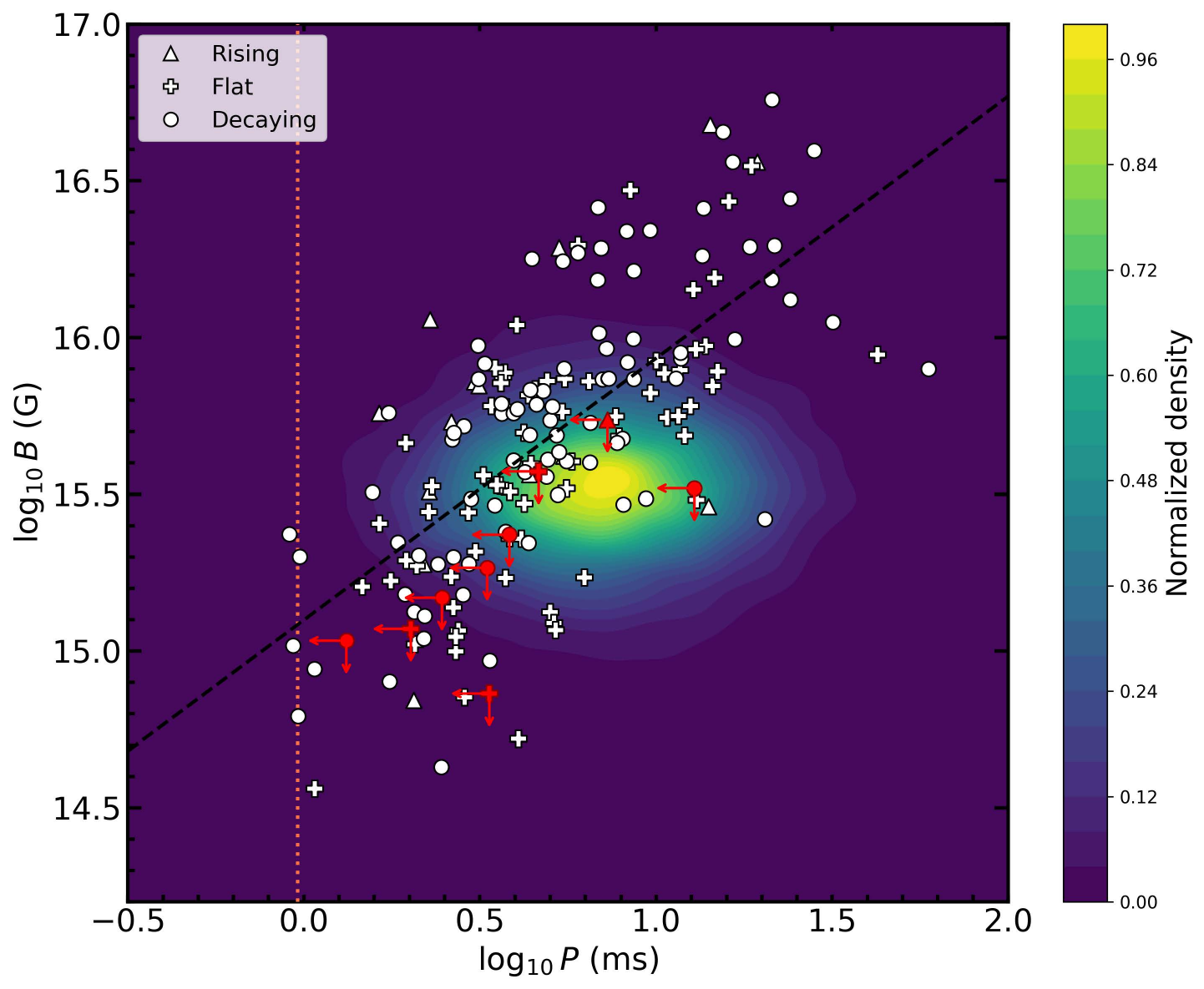}
\caption{Comparison of traditional inversion and Bayesian
population inference in the $B$--$P$ plane.
The colored background shows the Bayesian-inferred population
density, with the color bar indicating the normalized density.
The superimposed points show the traditional inversion results,
assuming $\eta=0.7$. Triangles, crosses, and circles denote
the rising, flat, and decaying plateau subsamples, respectively.
The red dots mark GRBs with internal plateaus;
downward and leftward arrows indicate upper limits on
$B_{\rm p}$ and $P_0$, respectively.
The black dashed line is the best-fitting power-law relation for
the traditional inversion results, excluding internal-plateau
events because they provide only upper limits.
The vertical dotted line marks the neutron-star break-up spin-period limit \citep{2004Sci...304..536L}.
}
\label{fig6}
\end{figure*}

Figure~\ref{fig6} compares the traditional magnetar estimates
with the Bayesian posterior distribution in the $B-P$ plane.
Despite methodological differences, the two approaches occupy
similar regions, with the Bayesian posterior forming a more
localized distribution largely enclosed by the direct estimates.
For the traditional inversion, we adopt $\eta=0.7$;
changing $\eta$ from 0.1 to 1 shifts individual
bursts in the $B-P$ plane but preserves the substantial overlap
between the two distributions. The larger spread of the direct
estimates reflects the event-by-event nature of the inversion,
whereas the Bayesian framework incorporates intrinsic scatter
and selection effects at the population level.

The rising, flat, and decaying plateau samples also show
substantial overlap in the $B-P$ plane and occupy the same broad
parameter space. This provides a complementary check on the CPP
inference that the observed diversity of plateau morphologies does
not require distinct magnetar populations and remains compatible
with a common underlying magnetar population.

GRBs with a plateau followed by an extremely steep decay
($\alpha_2\lesssim-3$) are commonly classified as internal
plateaus and are often associated with the collapse of a
supra-massive magnetar after it loses rotational support
\citep{2007ApJ...665..599T, 2010MNRAS.402..705L,
2017A&A...605A..60B}. In Figure~\ref{fig6}, these events appear
preferentially in the lower-left part of the inferred $B-P$
distribution. However, this apparent tendency should not be
interpreted as evidence for a distinct population, because the
inferred locations depend sensitively on the assumed efficiency,
neutron-star properties, and whether the observed break time is
interpreted as the dipole spin-down timescale or as a
collapse/shutoff time.

A power-law fit to the traditional inversion results yields a
slope of $0.84\pm0.06$, with a Pearson coefficient of $r=0.71$. A
comparable posterior correlation coefficient, $r\simeq0.73$, is
also obtained from the Bayesian population model. The
correlation is somewhat expected from the magnetic-dipole mapping
itself rather than reflecting an intrinsic correlation, as $B_{\rm
p}$ and $P_0$ are jointly inferred from the same plateau
luminosity and duration. The fitted slopes for the Rising, Flat,
and Decaying subsamples are $0.81\pm0.32$, $0.84\pm0.09$, and
$0.87\pm0.09$ respectively, which are mutually consistent within
their uncertainties. Positive $B$-$P$ correlations have been
interpreted in the context of fallback-accreting magnetar models,
where magnetosphere--disk interaction can drive the system toward
spin equilibrium \citep{2018ApJ...869..155S, 2026ApJ..1004L..11Z}.
However, the present correlation alone cannot uniquely establish
the operation of the propeller mechanism. 

A notable result is that the intrinsic luminosity scatter
consistently exceeds the duration scatter across all subsamples
and classification criteria. The smaller duration scatter suggests
a more stable plateau timescale, whereas the luminosity is more
variable across sources.
Variations in radiative efficiency can directly rescale the
plateau luminosity, while different jet opening angles
change the beaming correction to the isotropic-equivalent luminosity
\citep{2014ApJ...785...74L, 2014MNRAS.443.1779R}.
Viewing-angle effects may also broaden the observed luminosity distribution
in structured-jet models \citep{2020MNRAS.492.2847B}.
These effects are absorbed into the inferred intrinsic luminosity scatter
in our current framework, preventing us from determining which contribution dominates.

\subsection{Implications for Plateau Diversity}

Our CPP analysis indicates that the rising, flat, and decaying
plateau samples are compatible with a common underlying
magnetar population. The observed diversity of plateau
morphologies therefore need not imply distinct central-engine
populations. It may instead reflect differences in the
efficiency or environment through which energy
from a similar engine powers the afterglow.

In external-shock models, a broad range of
plateau morphologies can arise naturally from variations in
outflow properties \citep{2022NatCo..13.5611D, 2026MNRAS.549ag833S}, jet structure
\citep{2020MNRAS.492.2847B, 2020ApJ...893...88O}, and the
circumburst environment \citep{2020ApJ...903...18S,
2025ApJS..280...45L}.
Observational effects may further contribute to the apparent
diversity.
\citet{2009ApJ...690L.118Y} showed that some plateau features
can depend on the choice of reference time, while recent
large-sample studies found that the XRT observation start time
strongly influences plateau detectability and light-curve
complexity \citep{2025A&A...703A.101G, 2026A&A...710A.393V}.
These results suggest that several morphology-dependent
correlations can be weakened or even removed by observational
effects, and therefore should not be interpreted solely as
evidence for intrinsic differences in the central engine.

Pure magnetic-dipole spin-down cannot by itself produce a
pronounced rising plateau \citep{2001ApJ...552L..35Z,
2010MNRAS.402..705L}. The Rising subsample is therefore included
to test whether its characteristic luminosity and timescale imply
a magnetar parameter space distinct from the other morphologies.
Additional processes, such as fallback accretion, propeller
effects, jet structure, or magnetar-disc interactions, may modify
the energy injection and produce rising behaviors
\citep{2014MNRAS.438..240G, 2015MNRAS.446.3642Y,
2017MNRAS.470.4925G, 2018MNRAS.478.4323G}. For the Rising
subsample, the inferred $P$ and $B$ should therefore be regarded
as effective magnetar parameters within our dipole-based
framework, since additional processes may also affect their
inferred values. The broader Rising posteriors may reflect its
smaller sample size, larger physical diversity, or limitations of
the simple dipole model. Distinguishing these possibilities will
require larger samples and explicit morphology-dependent
modeling.

\section{Acknowledgements}
We thank the anonymous referee for the useful comments and
suggestions that led to an overall improvement of the
presentation. This study was supported by the National Natural
Science Foundation of China (Grant Nos. 12233002, 12273113,
12573051), by the National Key R\&D Program of China
(2021YFA0718500), by the Major Science and Technology Program of
Xinjiang Uygur Autonomous Region (No. 2022A03013-1). Y.F.H.
acknowledges the support from the Xinjiang Tianchi
Program.\vspace{5mm}

\section{Appendix}
\label{sec:Appendix}
\begin{figure*}
\centering
\includegraphics[width=0.95\textwidth]{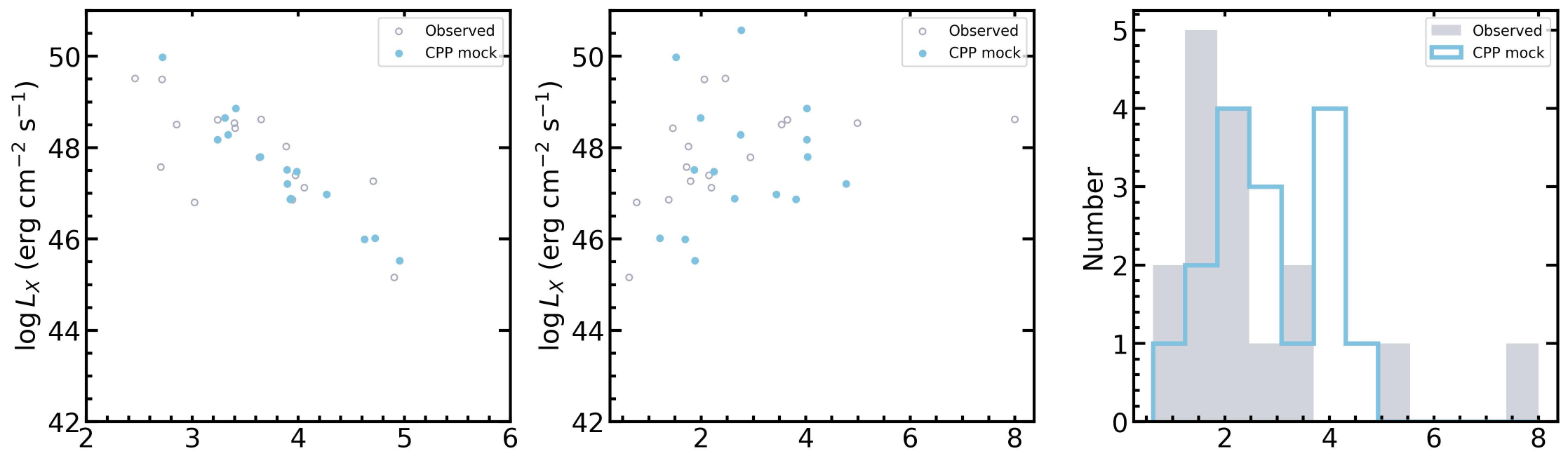}
\includegraphics[width=0.95\textwidth]{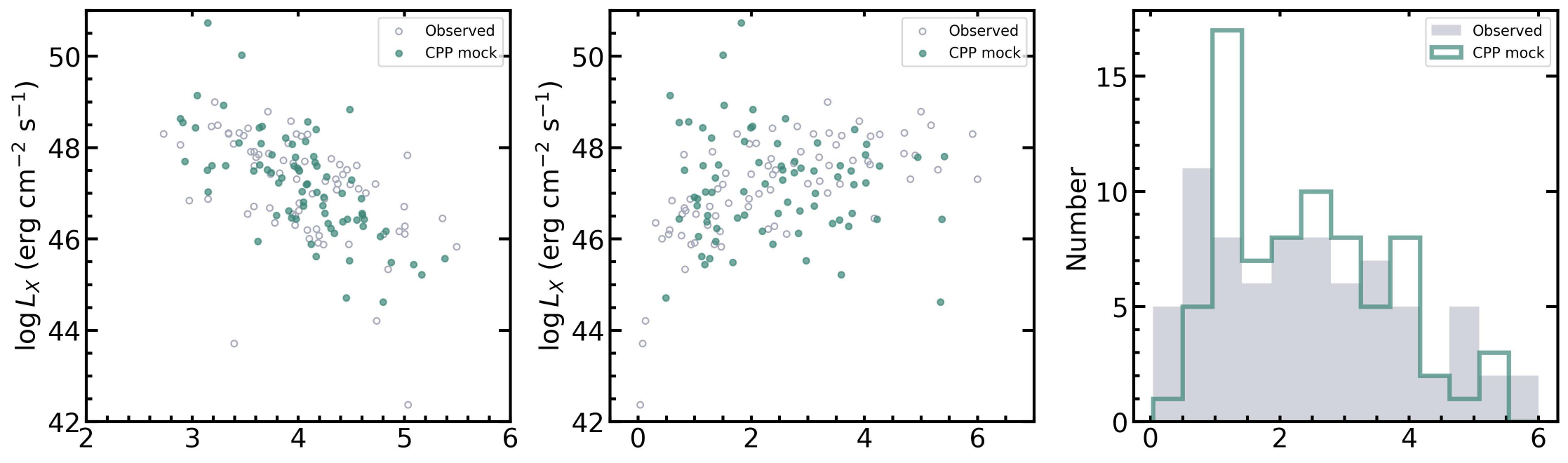}
\includegraphics[width=0.95\textwidth]{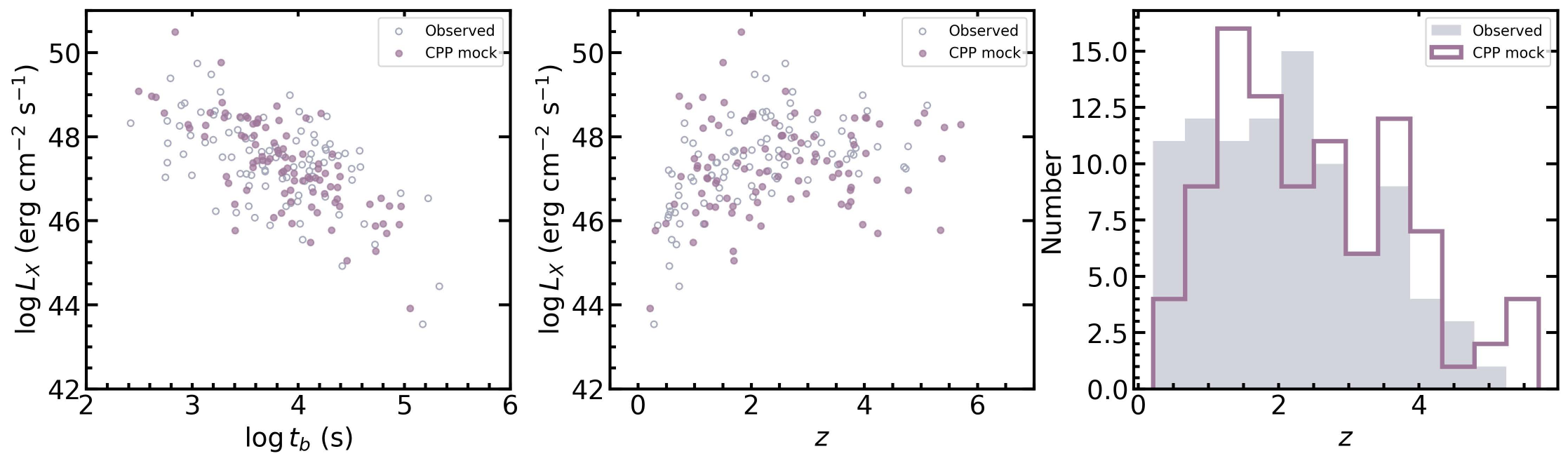}

\raggedright \textbf{Figure A1.} Comparison between the observed
and simulated distributions of \(L_{\rm X}\), \(t_{\rm b}\), and
\(z\) for the rising (the first row), flat (the second row), and
decaying (the last row) samples. The simulated distributions are
generated using the posterior-median parameter values. The three
columns show the $L_{\rm X}$--$t_{\rm b}$ plane, $L_{\rm X}$--$z$
plane, and redshift distribution, respectively.

\end{figure*}

\begin{figure*}

\label{fig:corner_all}
\centering
\includegraphics[width=0.49\textwidth]{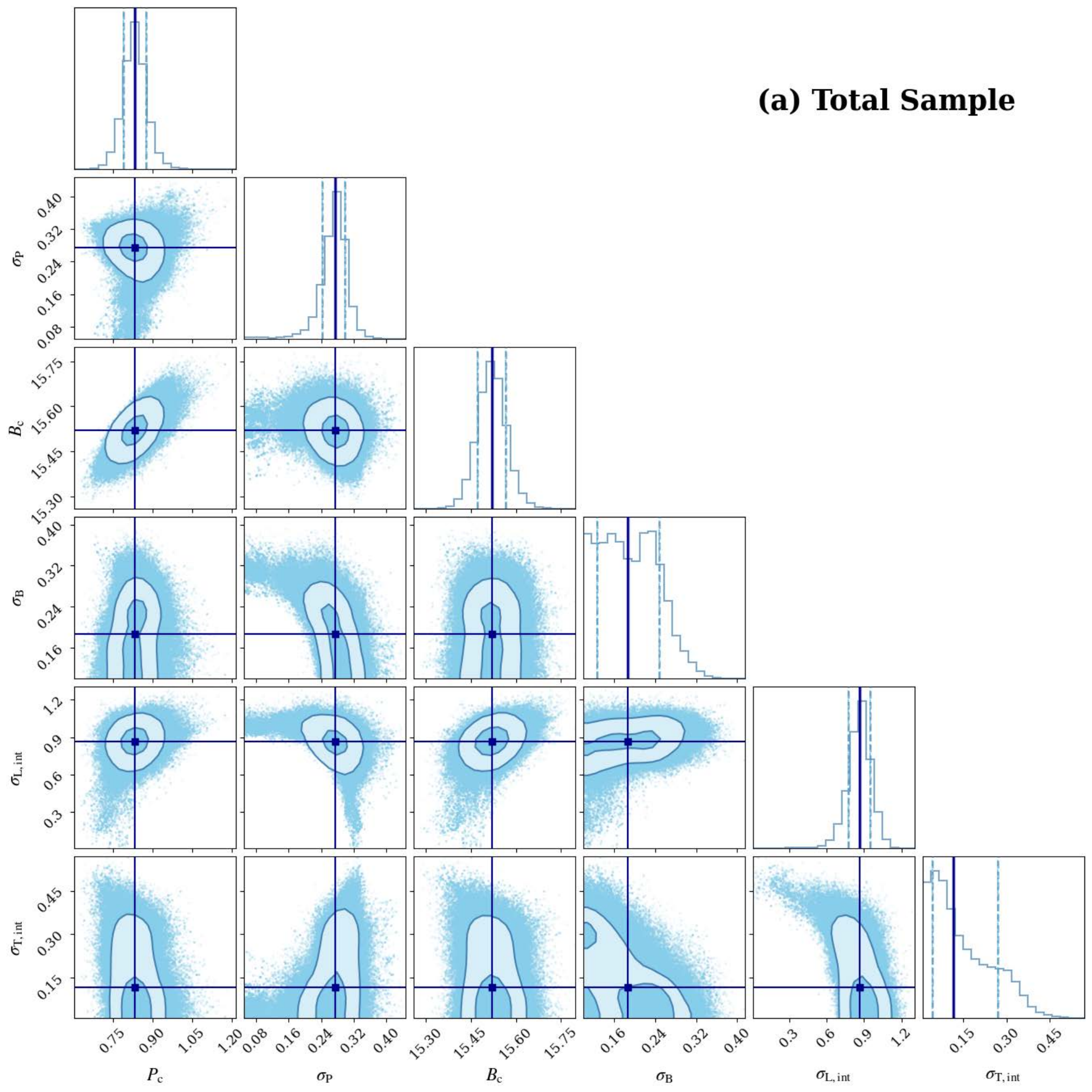}
\includegraphics[width=0.49\textwidth]{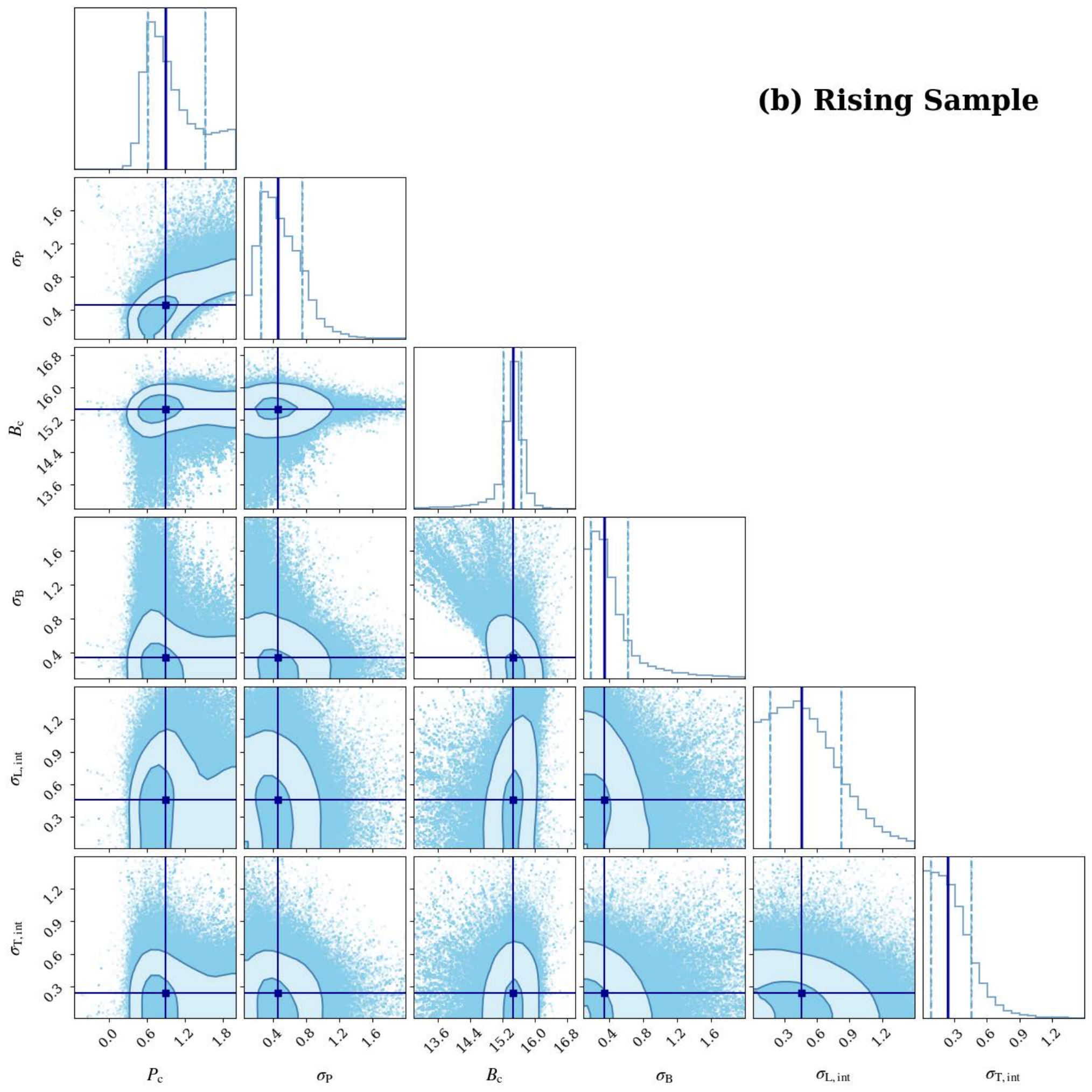}

\centering
\includegraphics[width=0.49\textwidth]{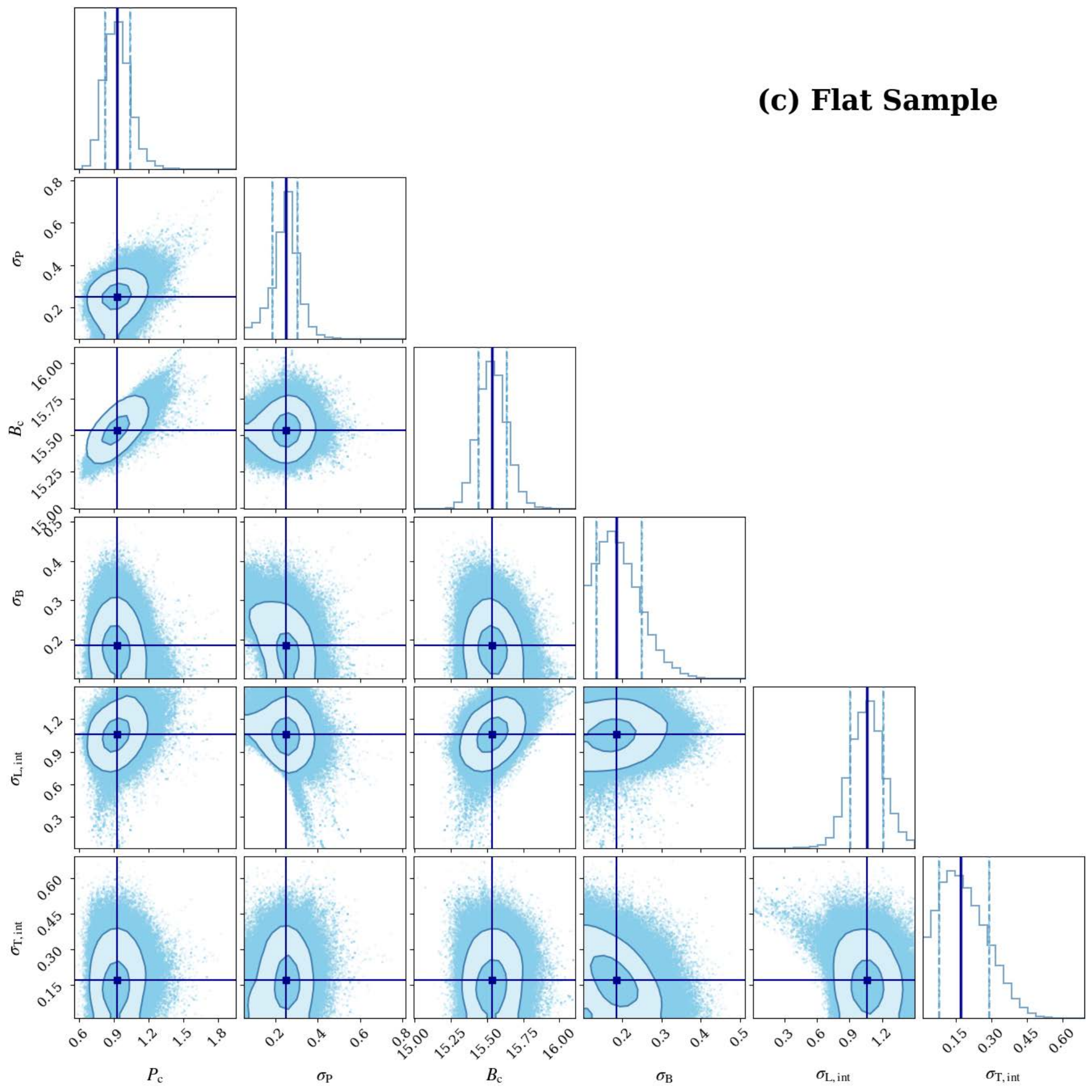}
\includegraphics[width=0.49\textwidth]{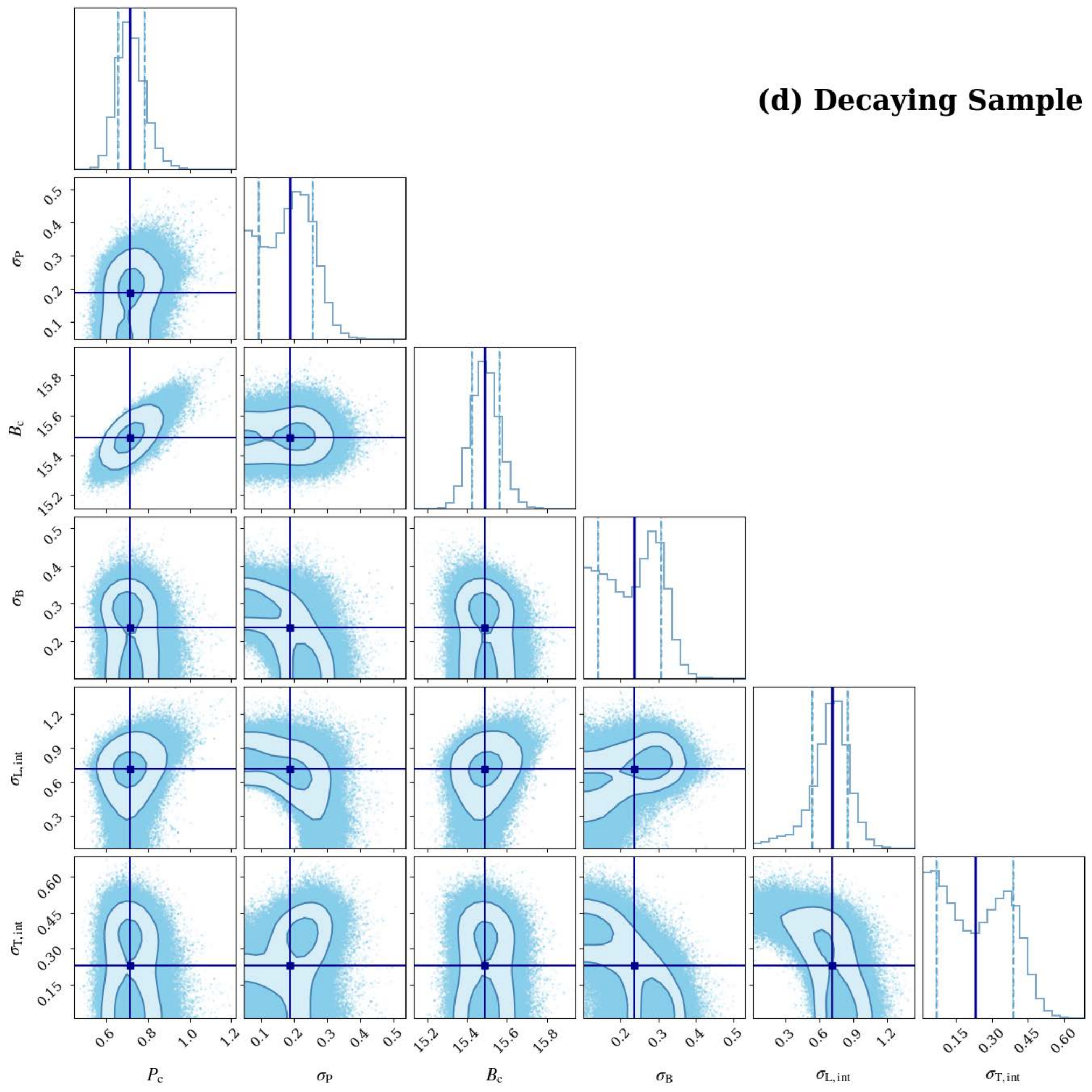}

\raggedright
\textbf{Figure A2.} Posterior distributions of the model
parameters inferred from the Markov chain Monte Carlo analysis.
The straight lines in the plot indicate the median
parameter values used to generate the simulated population.
\end{figure*}

\bibliography{main}{}
\bibliographystyle{aasjournal}

\end{document}